# Surface barrier of holes drilled in a type-II superconductor

D. M. Gokhfeld[1,2]

[1]Kirensky Institute of Physics, Krasnoyarsk Scientific Center, Siberian Branch, Russian Academy of Sciences, Krasnoyarsk, 660036 Russia.

[2]Siberian Federal University, Krasnoyarsk, 660041 Russia

**Abstract** Holes drilled in a type-II superconductor trap the magnetic flux. Following Clem's flux pinning model, we consider surface pinning as a mechanism for compressing the magnetic flux in the holes. Estimations of the trapped magnetic flux demonstrate that the holes with the diameter up to 2 mm are advantageous for bulk single-crystal REBCO samples. The REBCO films and tapes can be improved by the holes with diameter smaller than 10 μm.

Improving of magnetic and current-carrying properties of superconductors is a challenging and vital task for the sustainable growth [1]. One of the promising methods is perforation. It can improve oxygenation and prevent cracking of high-$T_c$ samples. Also, perforation directly affects the magnetic and current-carrying properties [2–7]. Holes can effectively trap magnetic flux [8]. In recent works [9, 10], where a flux trapping was modeled with using the Monte-Carlo method, values of the magnetic field in the holes were fitted to simulate well-defined peaks on the magnetic field profiles. The magnetic field trapped in the holes was used to be 20 times higher than the averaged local field.

In the presented study, the magnetic flux trapped in holes is found using Clem's surface pinning model [11]. It is proposed that the same barriers for Abrikosov vortices are raised by the superconductor surface and by holes drilled in the superconductor. Perforation parameters for the best values of trapped magnetic flux are estimated.

Consider a superconducting sample, the long cylinder with the diameter $D \gg \lambda$, where $\lambda$ is the London penetration length. The external magnetic field $H$ is applied parallel to the main axis of the cylinder, so the demagnetization is negligible. The sample contains holes which are drilled parallel to the main axis. It is assumed that the holes are uniformly distributed over the end face of the cylinder. The maximum value $H_{max}$ of $H$ is higher than the full penetration field $H_p$. The flux $\Phi$ trapped in the perforated sample after $H$ decreases to 0 is estimated using the critical state model [12]. Based on the result of the study [10], the following expression is proposed:

$$f = (1 - n_h)(1 + n_h k_h) \qquad (1)$$

where dimensionless parameters are used: $f$ is the reduced trapped flux, $f = \Phi/\Phi_{pin0}$, $n_h$ is the perforation coefficient, $n_h = A_h N_h S_h / S$, and $k_h$ is the reduced excess field in the holes, $k_h =$

$B_h/B_{pin0}$. The material-related ($\Phi_{pin0}$ is the flux trapped in the unperforated sample, $B_{pin0}$ is the corresponding remnant magnetic field, $B_{pin0} = \Phi_{pin0}/S$, and $S$ is the area of the sample end face) and the hole-related ($A_h$ is a coefficient depending on an arrangement of the holes, $N_h$ is the number of the holes, $S_h$ is the area of a single hole, and $B_h$ is the average magnetic field in the holes) parameters are involved. It should be noted that $A_h$ can be higher than 1 due to perturbations of current trajectories by the holes [10].

The expression (1) allows one to estimate the efficiency of pinning by holes and the perforation parameter for the maximum trapped flux. A condition for the increase of the trapped flux is $f > 1$. It is valid for

$$n_h < 1 - \frac{1}{k_h}. \tag{2}$$

The maximum value of trapped flux is

$$f = \frac{1}{4}\left(2 + k_h + \frac{1}{k_h}\right). \tag{3}$$

It is achieved for

$$n_h = \frac{1}{2}\left(1 - \frac{1}{k_h}\right). \tag{4}$$

Expressions (3-4) were obtained in the previous study [10]. However the value of $k_h$ was only a varied parameter where. Now we want to relate this parameter to the microscopic parameters of the superconductor. The holes are considered as convoluted surfaces into the superconductor. So an interaction of vortices with the superconducting surface is a subject of interest. The Clem surface pinning model [11] considers barriers for Abrikosov vortices appearing and disappearing near a surface of a type-II superconductor. The model statements are as follows: i) There is a vortex-free region near the surface. ii) The entry field $H_{en}$ determines the external field value required for an Abrikosov vortex to nucleate at the surface and move through the vortex-free region. ii) The exit field $H_{ex}$ needs for an Abrikosov vortex to pass back through the vortex-free region and leave the superconductor. Both $H_{en}$ and $H_{ex}$ depend on an averaged magnetic field $B$ at the inner border of the vortex free region:

$$H_{en} \approx \frac{1}{\mu_0}\left(B_s^2 + B^2\right)^{1/2},$$
$$H_{ex} \approx B/\mu_0, \tag{5}$$

where $B_s = \Phi_0/(4\pi\lambda\xi)$, $\xi$ is the coherence length, $\Phi_0 = 2.07*10^{-15}$ Wb, and $\mu_0 = 4\pi*10^{-7}$ H/m.

We assume that the hole has the similar surface barriers to vortices as the surface, and expression (5) is reliable to describe pinning of vortices in holes. In this case $H_{en}$ is a maximum value of the magnetic field trapped in the hole. Abrikosov vortices can enter to the holes only if the magnetic field in the hole is not higher than the averaged magnetic field $B$. Since $H_{en} > H_{ex}$, the holes facilitate the flux propagation but hinder the flux exit. The excess field in the holes is limited by the surface barrier. The reduced excess field is determined by $k_h \approx [(B_s/B_{pin0})^2+1]^{1/2}$.

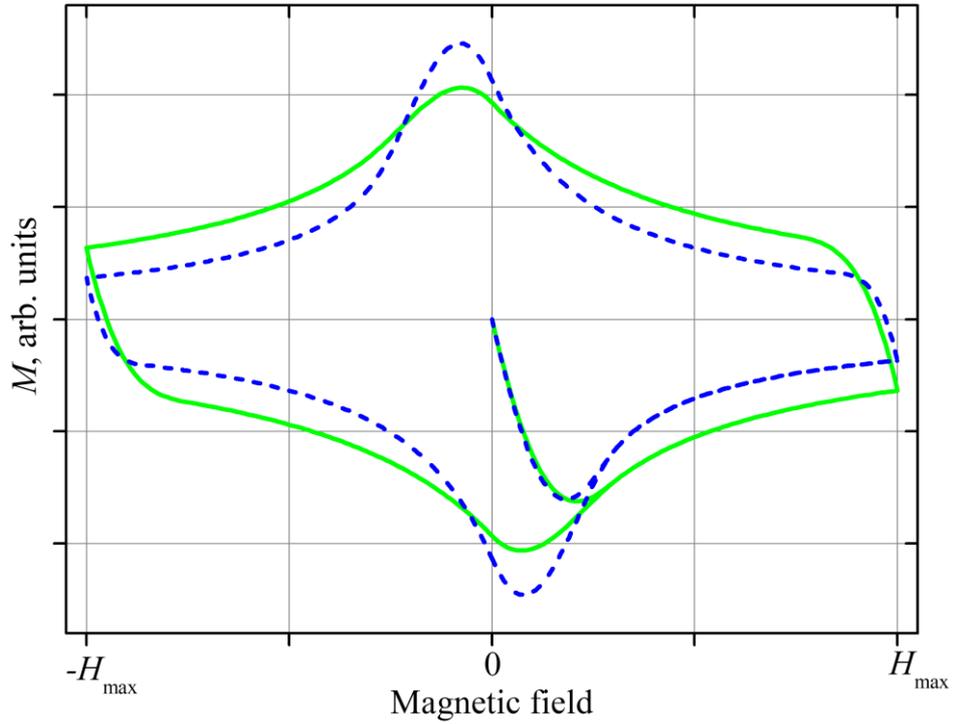

Fig. 1. Schematic representation of magnetization hysteresis loops for an unperforated sample (solid line) and a perforated sample (dashed line).

Mesoscopic samples may have $H_p < H_{en}$. This means that the condition $H_{max} > H_{en}$ may be wrong for these objects. In this case the magnetic field in the holes equals the average magnetic field in the sample at the external field $H_{max}$. One can estimate $k_h \approx (1/3)\mu_0 H_{max}/B_{pin0}$ for the considered cylindrical geometry.

The holes as well as the sample surface have the vortex-free region around their perimeter. The thickness of this region is not higher than $\lambda$ [11]. As the vortex-free region affects the reversible magnetization [13, 14], it is expected that the irreversibility field is reduced by perforation.

Any cavities in a type-II superconductor facilitate the propagation of magnetic flux inside. Thus, the perforated sample has smaller values of the penetration field $H_p$ and the magnetization width $\Delta M$ at $H > H_p$ than those of the unperforated sample. Then the external field is decreased, magnetic flux partially leaves the sample and the averaged magnetic field $B$ decreases. At the same time the holes prevent the magnetic flux from escaping. As a result, the values of $\Delta M$ at $H < H_p$ and the remnant magnetization $M_{rem} = B_{pin0}/\mu_0$ are higher for the perforated sample than for the unperforated one. Schematic representations of the magnetization hysteresis loops for perforated and unperforated samples are shown in Figure 1. The Kim-type dependence $j_c(B) = j_{c0}/(1 + B/B_1)$ is used to calculate these loops. The perforated sample is appointed to have a higher value of the zero-field critical current density $j_{c0}$ and a smaller value of the parameter $B_1$ than the unperforated sample.

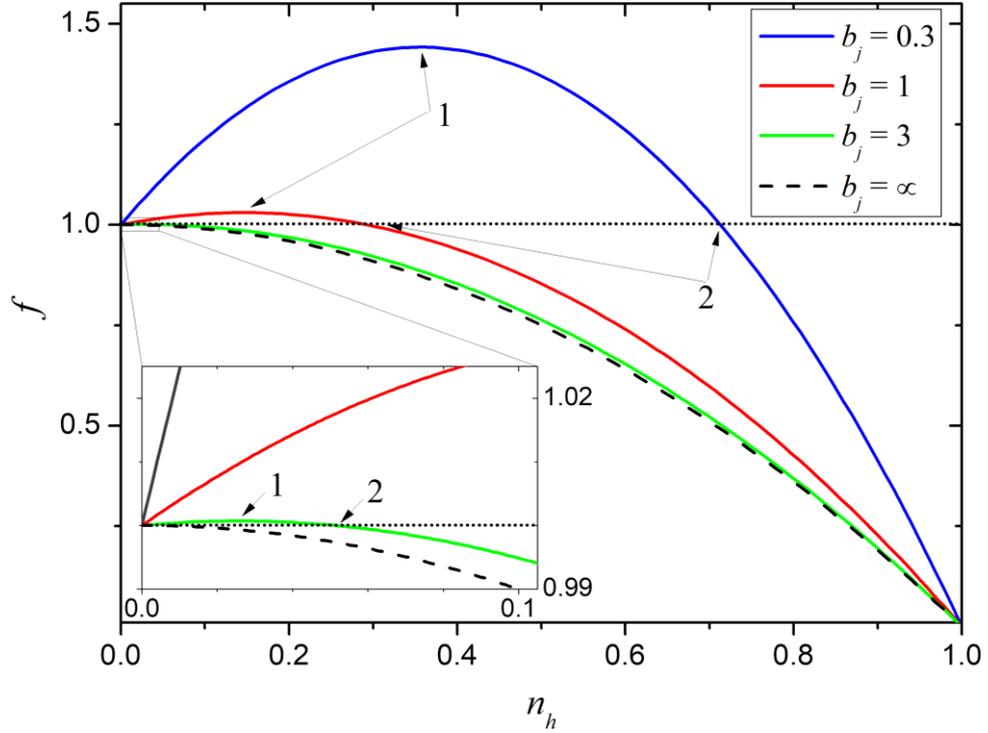

Fig. 2. Dependence of trapped flux $f$ on the perforation coefficient $n_h$ for different values of $b_j$. Inset demonstrates the same $f(n_h)$ curves at the interval of small $n_h$. Arrows indicate 1) the positions of maximal tapped flux and 2) the limiting values of $n_h$ for $f > 1$, i.e. for pinning advantage.

Next, the effect of the sample parameters on the trapped flux is analyzed. Let us introduce a material dependent parameter $b_j = B_{pin0}/B_s$. Figure 2 demonstrates the $f(n_h)$ dependence (1) calculated for different values of $b_j$. It can be seen that the $f(n_h)$ dependencies always have a maximum (points marked 1) for the finite values of $b_j$, and the maximum position shifts to smaller $n_h$ for higher $b_j$. Positions of $n_h$ corresponding to pinning advantage $f > 1$ (points marked 2) have the same behavior. We plot the positions of these points versus the material parameter $b_j$ (Fig. 3). These plots allow one to choose optimal perforation parameters for known values of $j_c$ for different superconducting materials.

It appears that the most experimental data of perforated high-$T_c$ samples can be divided into two groups with different $j_c$. Bulk REBCO (RE is Y or rare-earth elements) single-crystals have smaller values of $j_c$, the typical values are $j_c \sim 0.001 j_d$ at the liquid helium temperature and more than 10 times smaller at liquid nitrogen temperature [7, 15, 16], where $j_d$ is the depairing current density. REBCO films and tapes have higher values of $j_c$, they can have $j_c \sim 0.1 j_d$ at the liquid helium temperature [17, 18].

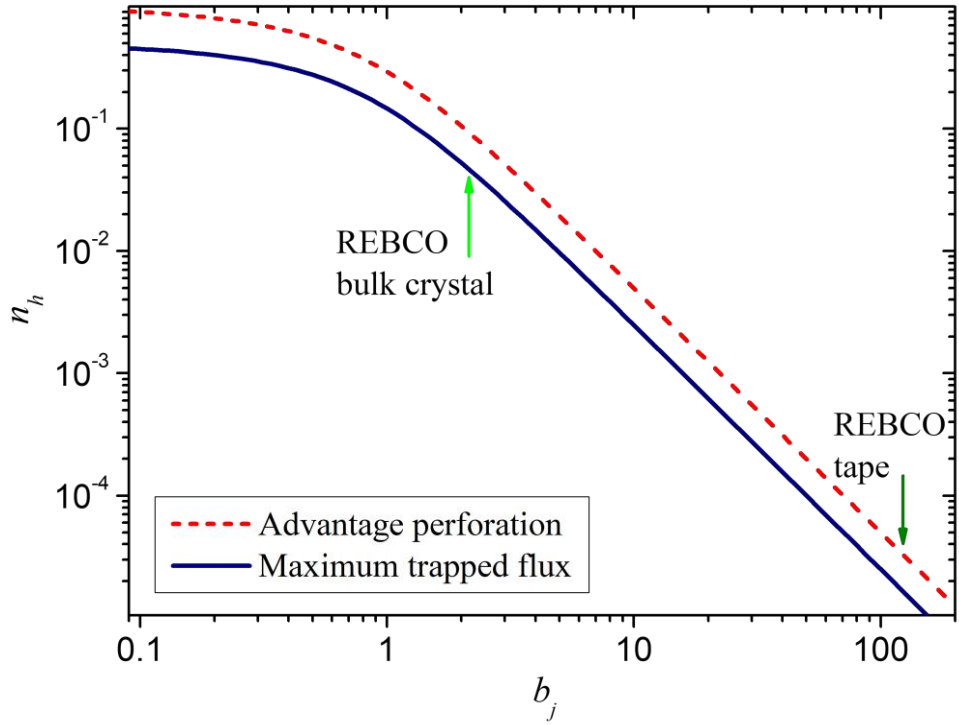

Fig. 3. The perforation coefficient $n_h$ providing pinning advantage (dash line) and maximum trapped flux (solid line) versus the material parameter $b_j$.

Using $B_{pin0} \approx \mu_0 j_c D/6$ for the considered cylinder geometry and given typical parameters of large-grain bulk REBCO samples ($D = 2$ cm, $j_c = 10^8$ A/m$^2$ and $B_s = 0.17$ T at 77 K), one obtains $b_j \approx 2.5$. In this case, the trapped flux enhancement requires $n_h < 0.07$. Optimal pinning can be achieved for 15 holes with $D_h = 1$ mm or for 60 holes with $D_h = 0.5$ mm. These holes are admitted to be suitable for the refrigerating liquid nitrogen. REBCO tapes with $j_c \approx 10^{10}$ A/m$^2$ and $D = 1$ cm have $b_j \approx 120$. This corresponds to $n_h < 3*10^{-5}$ and $D_h < 60$ μm. Optimal pinning can be achieved for about 1600 holes with $D_h = 1$ μm or for 16 holes with $D_h = 10$ μm. This means that the holes in such tapes cannot simultaneously provide the effective cooling and the pinning enhancement.

In conclusion, the surface barrier created by drilled holes in the superconducting sample is used to find the magnetic flux trapped in the holes. An influence of perforations on magnetization hysteresis loops of superconductors is considered. Perforation is expected to decrease the full penetration field and the irreversibility field. The remnant magnetization can be increased for some interval of the perforation parameter. The conditions for this increase are established. Optimal perforation parameters are found for REBCO bulk single crystals and REBCO tapes.

**References**

1. Muralidhar, M., Srikanth, A.S., Pinmangkorn, S., Santosh, M., Milos, J.: Role of Superconducting Materials in the Endeavor to Stop Climate Change and Reach Sustainable Development. J. Supercond. Nov. Magn. 1–10 (2023). https://doi.org/10.1007/s10948-023-06515-6
2. Bartolomé, E., Granados, X., Puig, T., Obrados, X., Reddy, E.S., Kracunovska, S.: Critical state of YBCO superconductors with artificially patterned holes. IEEE Trans. Appl. Supercond. 15, 2775–2778 (2005). https://doi.org/10.1109/TASC.2005.848210
3. Lousberg, G.P., Ausloos, M., Vanderbemden, P., Vanderheyden, B.: Bulk high-Tc superconductors with drilled holes: How to arrange the holes to maximize the trapped magnetic flux? Supercond. Sci. Technol. 21, 025010 (2008). https://doi.org/10.1088/0953-2048/21/02/025010
4. Jang, G., Lee, M., Han, S., Kim, C., Han, Y., Park, B.: Trapped field analysis of a high temperature superconducting bulk with artificial holes. J. Magn. 16, 181–185 (2011). https://doi.org/10.4283/JMAG.2011.16.2.181
5. Pannetier, M., Wijngaarden, R.J., Fløan, I., Rector, J., Dam, B., Griessen, R., Lahl, P., Wördenweber, R.: Unexpected fourfold symmetry in the resistivity of patterned superconductors. Phys. Rev. B. 67, 212501 (2003). https://doi.org/10.1103/PhysRevB.67.212501
6. Berdiyorov, G.R., Milošević, M. V., Peeters, F.M.: Composite vortex ordering in superconducting films with arrays of blind holes. New J. Phys. 11, 013025 (2009). https://doi.org/10.1088/1367-2630/11/1/013025
7. Kuchárová, V., Diko, P., Volochová, D., Antal, V., Lojka, M., Hlásek, T., Plecháček, V.: Microstructure and superconducting properties of bulk EuBCO-Ag with and without holes. J. Eur. Ceram. Soc. 42, 6533–6541 (2022). https://doi.org/10.1016/j.jeurceramsoc.2022.06.081
8. Matsumoto, K., Mele, P.: Artificial pinning center technology to enhance vortex pinning in YBCO coated conductors. Supercond. Sci. Technol. 23, 014001 (2010). https://doi.org/10.1088/0953-2048/23/1/014001
9. Maksimova, A.N., Kashurnikov, V.A., Moroz, A.N., Gokhfeld, D.M.: Trapped Field in Superconductors with Perforations. J. Supercond. Nov. Magn. 35, 283–290 (2022). https://doi.org/10.1007/s10948-021-06067-7

10. Gokhfeld, D.M., Maksimova, A.N., Kashurnikov, V.A., Moroz, A.N.: Optimizing trapped field in superconductors with perforations. Phys. C Supercond. its Appl. 600, 1354106 (2022). https://doi.org/10.1016/j.physc.2022.1354106

11. Clem, J.R.: A Model for Flux Pinning in Superconductors. In: Low Temperature Physics-LT 13. pp. 102–106. Springer US, Boston, MA (1974)

12. Bean, C.P.: Magnetization of high-field superconductors. Rev. Mod. Phys. 36, 31–39 (1964). https://doi.org/10.1103/RevModPhys.36.31

13. Burlachkov, L., Geshkenbein, V.B., Koshelev, A.E., Larkin, A.I., Vinokur, V.M.: Giant flux creep through surface barriers and the irreversibility line in high-temperature superconductors. Phys. Rev. B. 50, 16770–16773 (1994). https://doi.org/10.1103/PhysRevB.50.16770

14. Zeng, X.L., Karwoth, T., Koblischka, M.R., Hartmann, U., Gokhfeld, D., Chang, C., Hauet, T.: Analysis of magnetization loops of electrospun nonwoven superconducting fabrics. Phys. Rev. Mater. 1, 044802 (2017). https://doi.org/10.1103/PhysRevMaterials.1.044802

15. Lousberg, G.P., Fagnard, J.F., Haanappel, E., Chaud, X., Ausloos, M., Vanderheyden, B., Vanderbemden, P.: Pulsed-field magnetization of drilled bulk high-temperature superconductors: flux frontpropagation in the volume and on the surface. Supercond. Sci. Technol. 22, 125026 (2009). https://doi.org/10.1088/0953-2048/22/12/125026

16. Li, B., Zhou, D., Xu, K., Hara, S., Tsuzuki, K., Miki, M., Felder, B., Deng, Z., Izumi, M.: Materials process and applications of single grain (RE)–Ba–Cu–O bulk high-temperature superconductors. Phys. C Supercond. its Appl. 482, 50–57 (2012). https://doi.org/10.1016/j.physc.2012.04.026

17. George, J., Jones, A., Al-Qurainy, M., Fedoseev, S.A., Rosenfeld, A., Pan, A. V.: Tunable pinning effects produced by non-uniform antidot arrays in YBCO thin films. Ann. Phys. 529, 1600283 (2017). https://doi.org/10.1002/andp.201600283

18. Pokrovskii, S. V., Mavritskii, O.B., Egorov, A.N., Mineev, N.A., Timofeev, A.A., Rudnev, I.A.: Influence of ultrashort laser drilling on magnetic and transport characteristics of HTS tapes. Supercond. Sci. Technol. 32, 075008 (2019). https://doi.org/10.1088/1361-6668/ab14a3